# Compact, thermal-noise-limited optical cavity for diode laser stabilization at 1 x $10^{-15}$

A. D. Ludlow, X. Huang*, M. Notcutt, T. Zanon, S. M. Foreman, M. M. Boyd, S. Blatt, and J. Ye

JILA, National Institute of Standards and Technology and University of Colorado and Department of Physics, University of Colorado, Boulder, Colorado 80309-0440

Abstract

We demonstrate phase and frequency stabilization of a diode laser at the thermal noise limit of a passive optical cavity. The system is compact and exploits a cavity design that reduces vibration sensitivity. The sub-Hz laser is characterized by comparison to a second independent system with similar fractional frequency stability (1 x $10^{-15}$ at 1 s). The laser is further characterized by resolving a 2 Hz wide, ultranarrow optical clock transition in ultracold strontium.

OCIS codes: 140.2020, 030.1640, 300.6320

Highly frequency-stabilized lasers are essential in high resolution spectroscopy and quantum measurements [1], optical atomic clocks [2,3], and quantum information science [4]. To achieve superior stability with high bandwidth control, the laser output is traditionally servo locked to a highly isolated passive optical cavity using the Pound-Drever-Hall (PDH) technique [5]. Typical limits to the resulting laser stability include vibration-induced cavity length fluctuation, photon detection shot noise, and thermal-mechanical noise of the passive cavity components [6,7]. By thoughtful system design, the relative impact of these noise contributions can be adjusted. Here we report a cavity-laser system that achieves high stability with an appropriate compromise between acceleration sensitivity and thermal noise. The compact, cavity-stabilized diode laser has a sub-Hz linewidth (at several seconds) and is thermal-noise-limited to fractional instability of $\sim 1 \times 10^{-15}$ at time scales of 0.5 to 300 s. As further evidence of the laser's optical stability, we use it to interrogate an ultranarrow optical atomic transition in neutral atomic strontium. We resolve an optical transition linewidth of ~2 Hz, the narrowest optical atomic resonance observed to date.

The laser source in this work is a diode laser in an external cavity (ECDL) in the Littman configuration operating at 698 nm [Fig. 1(a)]. The laser is first pre-stabilized to a simple optical cavity with finesse of ~10,000. The PDH stabilization is accomplished via feedback to the laser diode current and the laser cavity piezo transducer (PZT). The servo bandwidth is 2–3 MHz. The prestabilized laser light is first-order diffracted by an acousto-optic modulator (AOM) and fiber-coupled to a platform on which an ultrastable cavity resides. This platform is mounted on a commercially available passive vibration isolation unit ([8], resonant frequency 0.5-1 Hz). Both the platform and the isolation unit are within a temperature-controlled enclosure lined with acoustic-damping foam. On the platform, the prestabilized laser light is phase modulated by an electro-optic modulator (EOM) operating at 5 MHz. Approximately 10 µW of optical power is incident on the ultrastable cavity for PDH locking. Stabilization to this cavity is accomplished via feedback to the AOM and a PZT controlling the prestabilization cavity length. The servo bandwidth for this final locking stage is ~100 kHz. The useful output of the laser is delivered via a fiber port located near the ultrastable cavity on the same platform. The entire optical set up occupies less than 1 m$^3$.

The ultrastable cavity has a finesse of 250,000 and is 7 cm long (cavity linewidth of ~7 kHz). Both the spacer and the optically bonded mirror substrates are made of ultralow expansion glass (ULE). To maintain small sensitivity of cavity length to acceleration, we implemented the following design features. First, because the acceleration sensitivity of fractional cavity length scales with the cavity length, we chose a somewhat short cavity spacer (7 cm) [9]. Second, the cavity is held at its midplane to achieve symmetric stretching and compressing of the two halves of the cavity during acceleration to suppress vibration sensitivity [9,10,11]. The cavity is mounted vertically to exploit the intuitive symmetry in the vertical direction [see Fig. 1(b)]. This mounting is accomplished by a monolithically attached midplane ring resting on three Teflon rods. Third, the cavity is wider in the middle and tapered at the ends, allowing more rigid construction without excess material. This cavity was designed in our laboratory and constructed in conjunction with a dozen other quantum metrology laboratories around the world. It is now commercially available, facilitating 1 Hz laser stabilization to any interested research lab [12]. The cavity and supporting rods are held in vacuum (10$^{-6}$ torr) by an ion pump (2 l/s); the cavity spacer has symmetric evacuation holes perpendicular to the optical axis. The vacuum can is single point temperature controlled to ~305 K within 500 μK over a 24 hour period. Since the ion pump is not temperature controlled, we installed a blackbody radiation baffle between the vacuum can and the ion pump.

To evaluate the final laser stability, a second cavity-laser system was constructed with a separate diode laser and a separate ultrastable cavity mounted on an independent vibration isolation platform in an independent enclosure. Light was transferred from one system to the other via optical fiber (employing fiber-phase-noise cancellation [13]) and a heterodyne beat between the two stabilized lasers was detected (Fig. 2 inset). Linear drifts of ≤ 1 Hz/s of the heterodyne beat were removed by applying a feedforward linear correction to reduce the drift to less than 50 mHz/s. At 300 mHz resolution bandwidth (RBW), the laser linewidth is 400 mHz wide (full width at half maximum). By reducing the RBW to 150 mHz, linewidths of 220 mHz can be observed with nonnegligible power in low-frequency noise sidebands. The fractional linewidth is below $1 \times 10^{-15}$. The stability of one of the lasers as measured by the fractional Allan deviation is also shown in Fig. 2. This measurement was taken directly by counting the heterodyne beat under linear drift cancellation. The theoretical estimate of the thermal-noise-limited stability is indicated as a solid line in the figure. This thermal noise limit has nearly negligible contribution from the ULE spacer itself, while the contribution from the ULE mirror substrates is approximately 1.5 times that from the dielectric high-reflective coating [6]. Except for a small noise bump at 100 s, the laser stability from 0.5 to 300 s coincides precisely with the modeled thermal noise limit.

Laser phase coherence was also observed via time domain measurements. The heterodyne beat at ~200 MHz between the two laser systems was mixed down to 15 Hz. This 15 Hz signal and its sine wave fit are shown in Fig. 3(a), with a linear chirp to account for the simple residual linear drift between the two lasers. The fit shows that the lasers remain phase coherent within 1 rad at the optical frequency of ~4 x $10^{14}$ Hz for a period > 2 s.

The sensitivity of the cavity length to accelerations was measured by shaking the cavity and observing the additional frequency noise present on the laser tightly locked to the cavity resonance. This was also an optical heterodyne measurement, with the system that was not shaken serving as the reference oscillator. The vertical acceleration sensitivity was measured to be 30 kHz/m/s$^2$. The horizontal acceleration sensitivity was 20 kHz/m/s$^2$ at 5 Hz; it dropped to 5 kHz/m/s$^2$ at 15 Hz because of mechanical isolation provided by the Teflon cavity mounting posts. The relatively short cavity used here constitutes a compromise between cavity acceleration sensitivity and the fractional thermal noise contributions (from the mirrors) to the cavity length stability. This compromise facilitates impressive diode laser stability at the $10^{-15}$ level with relatively straightforward vibration isolation. The difference in performance between this system and that of the highest recorded stability is consistent with the difference in cavity length that scales the fractional thermal noise [14,6].

To see this compromise more quantitatively, Fig. 3(b) shows the laser frequency noise spectrum. Below 5 Hz, the laser noise is dominated by thermal noise. Also shown is the laser noise contribution due to cavity acceleration—this is simply the measured acceleration noise spectrum on the vibration isolation platform scaled by the empirically determined acceleration sensitivity given above. This contribution to laser noise is dominated by vertical sensitivity because the vertical acceleration spectral density greatly exceeds the horizontal. The thermal noise contributes roughly a factor of 2 to 4 more than the acceleration noise. Consequently, the system could be further improved by using similar, but longer, mounted optical cavities together with fine tuning of the symmetrical rejection of acceleration sensitivity [9]. In this case, the combined thermal and acceleration noise contributions can be kept small enough for laser fractional frequency stability at the low side of the $10^{-16}$ decade.

To further characterize the laser coherence properties, we used one of the stable 698 nm lasers to probe an ultranarrow, doubly forbidden optical clock transition ($^1S_0$-$^3P_0$, natural linewidth ~1 mHz) in atomic strontium [15]. We probed a collection of laser-cooled Sr atoms trapped in an optical lattice at a temperature of 2 μK. Atom interrogation was performed in the Lamb-Dicke regime where both Doppler and recoil effects were eliminated. The recovered optical atomic transition is shown in Fig. 4. The linewidth of 2 Hz is the narrowest optical atomic transition observed to date (line quality factor $Q > 2 \times 10^{14}$). This linewidth corresponds closely to the Fourier limit given by the probe pulse duration (480 ms). Such a spectrum requires 20–30 s overall scanning time. The heterodyne beat between the two stabilized lasers integrated over 30 s reveals a single laser linewidth of ~2 Hz, suggesting that laser stability is the limiting factor in observing the narrow spectra. This atomic measurement provides a clean, independent confirmation of overall laser performance.

We have demonstrated a simple, compact, and ultrastable laser system useful for high precision metrology, quantum measurements, and quantum information applications, all of which require long atom-light coherence times. With the recent success of optical-lattice-based clocks in observing very high line Q with large atom numbers [15,16], the fundamental quantum projection noise of such systems can allow for 1 s instabilities at 1 x $10^{-17}$, achievable only if laser stability can further improve to permit observation of narrower atomic spectra. Consequently, future progress in this field requires pushing lasers to even higher stability. This requires improving the thermal noise limitation of passive optical cavities. As discussed in the literature [6,7,11], this can be done in three ways: (1) lower temperatures, (2) longer cavities, and (3) substrate and dielectric coating materials with lower mechanical loss. While technical challenges exist for any of these, the approach demonstrated here is compatible with each of these three routes to improvement.

The authors thank J. Hall of JILA for his long-term interest and leadership in laser stabilization. We also thank T. Zelevinsky of JILA and R. Lalezari of Advanced Thin Films. This work is supported by ONR, NIST, and NSF.
* State Key Laboratory of Magnetic Resonance and Atomic an Molecular Physics, Wuhan Institute of Physics and Mathematics, Chinese Academy of Sciences, Wuhan, 430071, China
** Mention of commercial products is for information only; it does not imply NIST endorsement.

References (with titles)

1. R. J. Rafac, B. C. Young, J. A. Beall, W. M. Itano, D. J. Wineland, and J. C. Bergquist, "Sub-dekahertz Ultraviolet Spectroscopy of $^{199}Hg^+$," Phys. Rev. Lett. **85** 2462 (2000).

2. A. D. Ludlow, M. M. Boyd, T. Zelevinsky, S. M. Foreman, S. Blatt, M. Notcutt, T. Ido, and J. Ye, "Systematic Study of the $^{87}$Sr Clock Transition in an Optical Lattice," Phys. Rev. Lett. **96** 033003 (2006).

3. H. Stoehr, F. Mensing, J. Helmcke, and U. Sterr, "Diode laser with a 1 Hz linewidth," Opt. Lett. **31** 736 (2006).

4. F. Schmidt-Kaler, S. Gulde, M. Riebe, T. Deuschle, A. Kreuter, G. Lancaster, C. Becher, J. Eschner, H. Haffner, and R. Blatt, "The coherence of qubits based on single Ca+ ions," J. Phys. B **36** 623 (2003).

5. R. W. P. Drever, J. L. Hall, F. V. Kowalski, J. Hough, G. M. Ford, A. J. Munley, and H. Ward, "Laser Phase and Frequency Stabilization Using an Optical Resonator," App. Phys. B **31**, 97 (1983).

6. K. Numata, A. Kemery, and J. Camp, "Thermal-Noise Limit in the Frequency Stabilization of Lasers with Rigid Cavities," Phys Rev. Lett. **93**, 250602 (2004).

7. M. Notcutt, L.-S. Ma, A. D. Ludlow, S. M. Foreman, J. Ye, and J. L. Hall, "Contribution of thermal noise to frequency stability of rigid optical cavity via Hertz-linewidth lasers," Phys. Rev. A **73** 031804R (2006).

8. MinusK Inc **.

Fig. 1: (a) Schematic of the laser and passive cavity optical set up. EOM: electro-optic modulator; AOM: acousto-optic modulator; ¼: quarter waveplate; ½: half waveplate; isol: optical isolator; pbs: polarizing beam splitter; pd: photodetector. (b) The high-finesse, ultrastable ULE optical cavity in its vertical mounting configuration.

Fig. 2: Fractional Allan deviation of the stabilized laser. The solid line near a fractional frequency stability of $1 \times 10^{-15}$ denotes the thermal noise stability limit of the passive optical cavity. Inset: A measurement of laser linewidth from the heterodyne beat.

Fig. 3: (a) Heterodyne beat of the two stabilized lasers, mixed down to 15 Hz. The dots indicate the measured data and the solid curve is the chirped sine wave fit. The fit shows that the lasers maintain phase coherence within 1 radian for > 2 s. (b) Frequency noise spectrum of the stabilized laser (solid data). The solid line is the theoretical estimate of the thermal noise contribution, with its characteristic $1/\sqrt{f}$ dependence. The short-dashed data indicates contribution to laser frequency noise from the acceleration sensitivity of the optical cavity. The horizontal dashed line denotes the measurement noise floor and dominates laser frequency noise well above 10 Hz.

Fig. 4: Narrow atomic resonance observed by probing ultracold strontium with the stabilized laser. The 2 Hz full width at half maximum is the narrowest optical atomic resonance observed to date. The transition frequency, $f_0$, is ~$4.29 \times 10^{14}$ Hz.

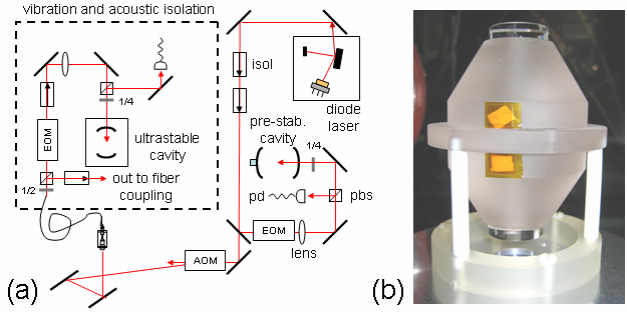

**Fig .1**

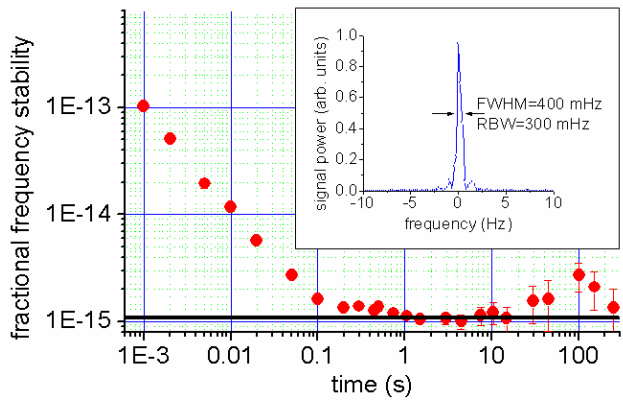

**Fig. 2**

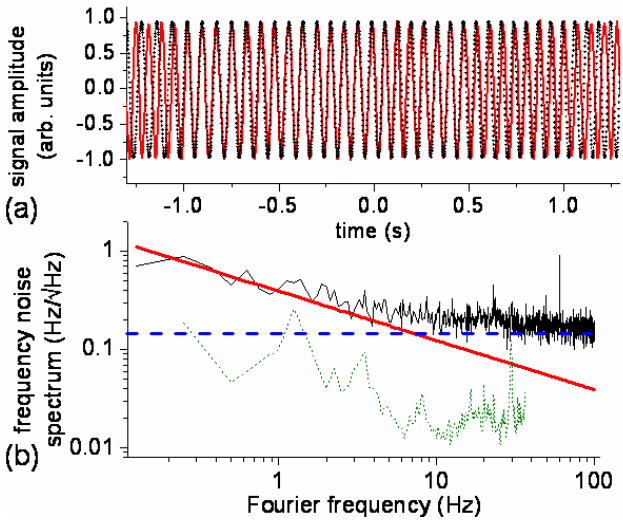

**Fig. 3**

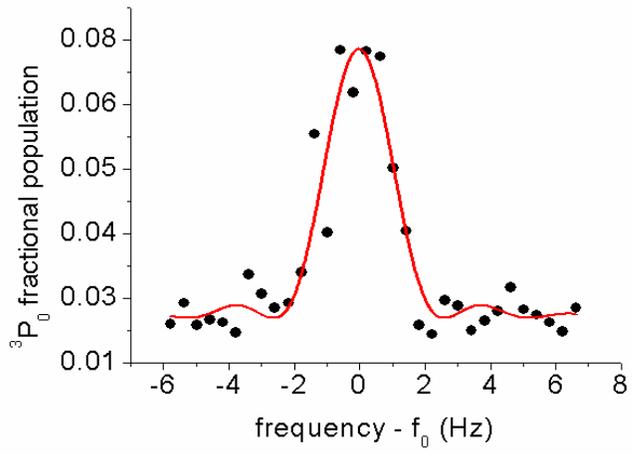

**Fig. 4**